\documentclass[superscriptaddress,twocolumn,prm,showkeys,showpacs,aps]{revtex4-2}
\usepackage[utf8]{inputenc}
\usepackage{amsmath}
\usepackage{amssymb}
\usepackage{amsfonts}     
\usepackage{graphicx}   
\usepackage{verbatim}  
\usepackage{color}     
\usepackage{subfigure}  
\usepackage{hyperref}   
\usepackage{float}
\usepackage{natbib}
\usepackage{gensymb}
\usepackage{color}
\usepackage{epstopdf}
\usepackage{multirow}
\usepackage{ulem}
\newcommand{\CM}[1]{\textcolor{black}{{#1}}}

\newif\ifCMbool
\CMboolfalse
\ifCMbool
\newcommand{\CMs}[1]{\textcolor{red}{{\sout{#1}}}}
\else
\newcommand{\CMs}[1]{\textcolor{red}{{ }}}
\fi

\begin{document}

\title{Insensitivity of the striped charge-orders in IrTe$_2$ to alkali surface doping implies their structural origin}

\author{M.~Rumo}
\altaffiliation{Corresponding author:\\ maxime.rumo@unifr.ch}
\affiliation{D{\'e}partement de Physique and Fribourg Center for Nanomaterials, Universit{\'e} de Fribourg, CH-1700 Fribourg, Switzerland}

\author{A.~Pulkkinen}
\affiliation{D{\'e}partement de Physique and Fribourg Center for Nanomaterials, Universit{\'e} de Fribourg, CH-1700 Fribourg, Switzerland}
\affiliation{School of Engineering Science, LUT University, FI-53850, Lappeenranta, Finland}

\author{B.~Salzmann}
\affiliation{D{\'e}partement de Physique and Fribourg Center for Nanomaterials, Universit{\'e} de Fribourg, CH-1700 Fribourg, Switzerland}

\author{G.~Kremer}
\affiliation{D{\'e}partement de Physique and Fribourg Center for Nanomaterials, Universit{\'e} de Fribourg, CH-1700 Fribourg, Switzerland}

\author{B.~Hildebrand}
\affiliation{D{\'e}partement de Physique and Fribourg Center for Nanomaterials, Universit{\'e} de Fribourg, CH-1700 Fribourg, Switzerland}

\author{K.Y.~Ma}
\affiliation{Department of Chemistry, University of Zurich, CH-8057 Zurich, Switzerland}

\author{F.O.~von~Rohr}
\affiliation{Department of Chemistry, University of Zurich, CH-8057 Zurich, Switzerland}

\author{C.W.~Nicholson}
\affiliation{D{\'e}partement de Physique and Fribourg Center for Nanomaterials, Universit{\'e} de Fribourg, CH-1700 Fribourg, Switzerland}

\author{T.~Jaouen}
\affiliation{Univ Rennes, CNRS, Institut de Physique de Rennes - UMR 6251, F-35000 Rennes, France}

\author{C.~Monney}
\altaffiliation{Corresponding author:\\ claude.monney@unifr.ch}
\affiliation{D{\'e}partement de Physique and Fribourg Center for Nanomaterials, Universit{\'e} de Fribourg, CH-1700 Fribourg, Switzerland}

\begin{abstract}
We present a combined angle-resolved photoemission spectroscopy and low-energy electron diffraction (LEED) study of the prominent transition metal dichalcogenide IrTe$_2$ upon potassium (K) deposition on its surface. Pristine IrTe$_2$ undergoes a series of charge-ordered phase transitions below room temperature that are characterized by the formation of stripes of Ir dimers of different periodicities. Supported by density functional theory calculations, we first show that the K atoms dope the topmost IrTe$_2$ layer with electrons, therefore strongly decreasing the work function and shifting only the electronic surface states towards higher binding energy. We then follow the evolution of its electronic structure as a function of temperature across the charge-ordered phase transitions and observe that their critical temperatures are unchanged for K coverages of $0.13$ and $0.21$~monolayer (ML). Using LEED, we also confirm that the periodicity of the related stripe phases is unaffected by the K doping. We \CM{surmise}\CMs{conclude} that the charge-ordered phase transitions of IrTe$_2$ are robust against electron surface doping, because of its metallic nature at all temperatures, and due to the importance of structural effects in stabilizing charge order in IrTe$_2$. 
\end{abstract}
\date{\today}
\maketitle

\section{Introduction}

Low-dimensional transition metal dicalcogenides (TMDCs) are very attractive for study because high quality crystals of large size can be grown and exfoliated down to the monolayer (ML). They display a wide range of physical properties and complex phase diagrams including charge-density waves (CDWs) and superconductivity~\cite{RossnagelIOPCDW,JohannesPRB,PyonJPSJ,MorosanNat,SiposNatMat,YangNatPhys}. Many of these compounds have a relatively simple low-energy electronic structure with a few relevant bands and can be seen as model systems for small gap semiconductors~\cite{ChernikovPRL}, valleytronics~\cite{MakNatNano,ZengNatNano,RileyNatPhys,BertoniPRL} and also topological properties~\cite{TamaiPRX,WuPRB,KingNatMat}. They have often been investigated by means of external perturbations, not only to understand their physical properties, but also to control them~\cite{Stojchevska1Science,OikeSuper,SieNat,NicholsonComMater2021}. In this framework, \textit{in situ} alkali deposition is an efficient and simple way to tune, suppress or generate new ground states. It has been shown to have a strong impact on most CDW phases in TMDCs~\cite{RossnagelIOPTiSe,jaouenarxiv}, to induce a surface semiconductor-semimetal transition in black phosphorus~\cite{KimScience} or even to enhance the excitonic insulator phase in Ta$_2$NiSe$_5$~\cite{FukutaniPRL}. To date, there is no such study on IrTe$_2$, an enigmatic large spin-orbit coupling TMDC exhibiting a complex succession of charge-ordered phases as a function of temperature~\cite{PascutPRB,MauererPRB,HsuPRL,RumoPRB}.

IrTe$_2$ undergoes several structural first-order phase transitions below room temperature (RT). The system goes from a trigonal unit cell of CdI$_2$-type ($P\overline{3}m1$) to a monoclinic unit cell ($P\overline{1}$) accompanied by a sudden jump in resistivity and magnetic susceptibility at $T_{c_1} = 278$~K~\cite{KoNatCom,FangScienRep,JobicZeit,MatsumotoJLTP,ToriyamaJapanLetters,KoleySSCom,LiSciRep,ParisXAS,RumoPRB}. In this first charge-ordered phase, one-dimensional stripes of Ir dimers~\cite{HsuPRL,PascutPRL} appear due to a large decrease of their bond length and lead to a $(5\times1\times5)$ superstructure~\cite{PascutPRL,PascutPRB,KoNatCom,LiNatCom,ToriyamaJapanLetters,OhPRL,TakuboPRB}. Although the changes of the in-plane bonding suggest a multi-center bond as a more complete description~\cite{SalehEntropy}, for brevity, we will continue to use “dimers” throughout the text. A second phase transition occurs at $T_{c_2} = 180$~K, characterized by a $(8\times1\times8)$ superstructure. This has stimulated numerous scanning tunneling microscopy (STM)~\cite{MauererPRB,KoNatCom,LiNatCom,DaiPRB} and angle-resolved photoemission spectroscopy (ARPES) studies~\cite{RumoPRB,LeeIOP,OotsukiJapanLettersES,OotsukiJournPhys,KoNatCom,NicholsonComMater2021}, which revealed additional periodicities and a surface periodicity $(6\times1)$ relative to the ground state reconstruction appearing after a third phase transition at $T_{c_3} = 165$~K. At low temperatures these phases coexist on the nanometer scale~\cite{HsuPRL} therefore complicating the interpretation of ARPES data which averages over the beam area. Substantial changes to the material behavior are produced by doping: superconductivity is induced by partial substitution of Ir with Pt~\cite{PyonJPSJ} or Pd~\cite{YangPRL}, or by temperature quenching~\cite{OikeSuper}, while partial substitution of Te with Se induces charge order~\cite{PascutPRB,DaiPRB,OhPRL}, further emphasizing the metastable nature of the material. Recently it has been shown that the application of uniaxial strain to IrTe$_2$ grants access to macroscopic regions of the $(6\times1)$ ground state and its corresponding topological states~\cite{NicholsonComMater2021}. This stabilization was found to be enabled by the strain-induced charge transfer from Ir to the out-of-plane antibonding Te orbitals, which modifies the energetic landscape of the competing phases. An open question is how the occurrence and periodicity of these different charge-ordered phases reacts to deposition of alkali atoms at the surface of IrTe$_2$, and whether a similar stabilizing effect as observed with strain may be achieved. 
\\

We report here on the effect of \textit{in situ} potassium (K) deposited at RT on the electronic structure of IrTe$_2$ by means of ARPES. Combined with density functional theory (DFT) calculations, we establish that the K atoms give most of their electronic charge to the surface layer, resulting in a significant decrease of the work function. As the doping density increases, the surface electronic states are progressively lowered towards higher binding energies (BE). Although the K atoms modify the surface electronic structure, temperature-dependent ARPES shows that the critical temperatures of the phase transitions remain unaffected, while low-energy electron diffraction (LEED) measurements of the low-temperature charge-ordered phases confirm that the stripes periodicities remain unchanged. This demonstrates that alkali doping, which electronically populates states close to the Fermi level, is not sufficient to interfere with the bonding-antibonding molecular states relevant for the phase transition that occur far from the Fermi energy. Despite the fact that alkali doping affects only the surface, we observe that the surface reconstruction is similarly unchanged. This suggests that local lattice effects are central to understand the formation of charge-ordered phase transitions in IrTe$_2$.

\section{Methods}

Single crystals of IrTe$_2$ were grown using the self-flux method~\cite{FangScienRep,JobicZeit}. They were characterized by magnetic susceptibility and resistivity measurements, which confirm that $T_{c_1}$ $= 278$~K and $T_{c_2}$ = $180$~K~\cite{RumoPRB}. Samples were cleaved at RT in vacuum at a pressure of about $10^{-8}$~mbar. During the photoemission measurements, the base pressure was better than $5\times10^{-11}$~mbar. K deposition was achieved \textit{in situ} by evaporation from a commercial SAES getter source in pressure below $5\times10^{-10}$~mbar. The temperature-dependent ARPES study was carried out using a Scienta DA$30$ photoelectron analyzer and two different excitation sources, namely monochromatized He$_{\text{I}}$ radiation ($h\nu=21.22$~eV) and a high-energy-resolution laser commercial setup (Harmonix, APE GmbH) generating $6.3$~eV photons using harmonic generation from the output of an optical parametric oscillator pumped by a Paladin laser (Coherent, inc.) at $80$~MHz. The total energy resolutions were about $5$ and $3$~meV at $21.2$ and $6.3$~eV photon energy, respectively, and the error on the sample temperature was estimated to be $5$~K. Cooling of the sample was carried out at rates $<3$ K$/$min and each measurement was preceded by a pause of at least 10 min, to ensure thermalization. Scanning tunneling microscopy (STM) measurements were performed on a commercial low-temperature STM (Scienta-Omicron) at $4.5$~K in fixed current mode and with a bias voltage applied to the sample. The LEED patterns were recorded with a SPECS ErLEED at $64$~eV.

DFT calculations with spin-orbit interaction were performed using the Vienna \textit{ab initio} simulation package (VASP)~\cite{kresse1993,kresse1994,kresse1996a,kresse1996b} within the projector augmented wave method~\cite{kresse1999} and the Perdew-Burke-Ernzerhof (PBE) functional~\cite{perdew1996}. The surface was modeled by a slab of three layers, with the bottom layer atoms fixed at bulk positions and the rest of the atom positions relaxed until the forces are $< 1$~meV/\AA. The K atoms were placed on the topmost surface, and dipole corrections were included to account for the adsorbate-induced dipole moment. The cutoff energy was set to $400$~eV and the $k-$point grid spacing was $<0.025$~\AA$^{-1}$. Band unfolding has been performed using the \textsc{PyProcar} code~\cite{herath2020}.

\section{Results}

\subsection{Potassium deposition characterization}

We first address the influence of K deposition on the surface electronic structure of IrTe$_2$ at RT. Combining ARPES and STM measurements on the sample, we are able to estimate the K coverage as a function of exposure time to the K evaporation source. More details are available in the Appendix~\ref{Potassium deposition calibration}. 

\begin{figure}[t]
\includegraphics[width=0.85\columnwidth]{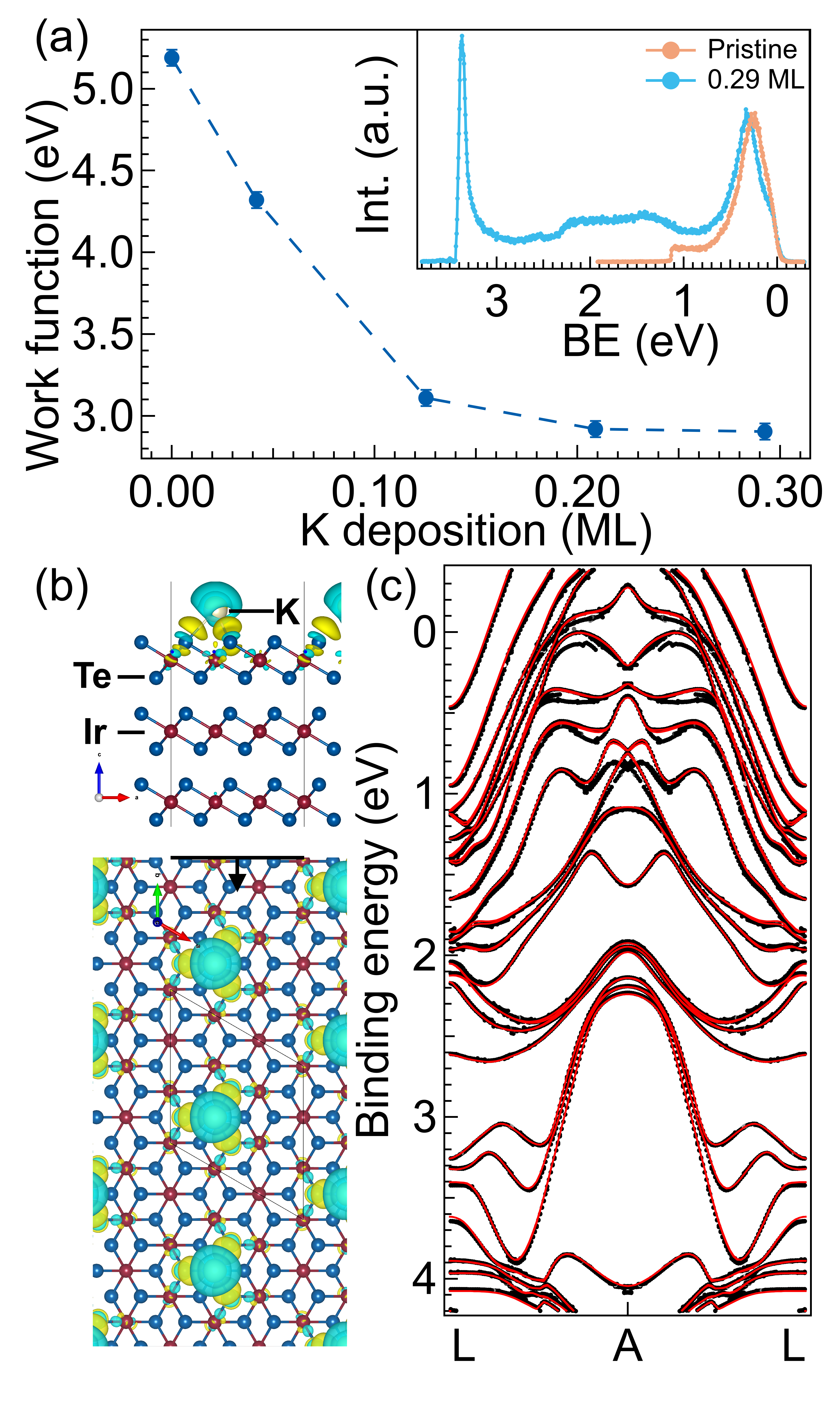}
\caption{\label{Figure WF} 
(Color online)~(a) Evolution of the work function of IrTe$_2$ as a function of K coverage. The inset shows energy distribution curves (EDCs) measured at RT with a photon energy of $h\nu = 6.3$~eV on a pristine crystal and $0.29$ ML K-doped crystal along the AL direction. (b) Side and top views of the IrTe$_2$ atomic structure with a $(3\times3)$ K adlayer on the surface. The light blue isosurface refers to the missing charge and the yellow isosurface to the gained charge at $0.001$~$e^{-}/a_{0}^3$~\cite{VESTA}, where $a_0$ is the Bohr radius. (c) Band structure calculation of the pristine IrTe$_2$ (red) and with a K adlayer as shown in graph (b) (black).} 
\end{figure}

Figure~\ref{Figure WF}~(a) displays the evolution of the work function of IrTe$_2$ with K deposition. The work function $\phi$, defined as the energy of the vacuum level $E_\text{vac}$ with respect to the Fermi level $E_{\text{F}}$, is determined from the low-energy cutoff $E_\text{cut}$ of the secondary photoelectron emission measured using 6.3 eV photons (see the inset), $\phi = h\nu- (E_F-E_\text{cut})$. With increasing K coverage, the work function decreases, and exhibits a saturation after a deposition of $0.2$ monolayer (ML) (1 ML corresponds to one K atom per IrTe$_2$ surface unit cell). For coverages below $0.13$~ML, a constant decrease of the work function is observed. In the inset are shown angle-integrated ARPES spectra of pristine IrTe$_2$ as well as IrTe$_2$ with the highest K doping used in this study. We have then performed DFT calculations to anticipate the influence of K deposition on the low energy surface electronic band structure of IrTe$_2$. Figure~\ref{Figure WF}~(b) displays the atomic structure of IrTe$_2$ with one K adatom on a $(3\times3)$ surface unit cell for a three-layers slab, corresponding to a $0.11$~ML coverage. The structure has been relaxed and the optimal K adsorption site has been found to be on top of an Ir atom. The changes to the charge distribution induced by the K adatom are also shown on the same image. Our calculations demonstrate that only the surface charge distribution is modified by the K adsorption, and according to Bader analysis~\cite{tang_2009,sanville_2007,henkelman_2006,yu_2011}, each K adatom gives about $0.63$~$e^{-}$ to the first IrTe$_2$ surface layer, corresponding to $0.07$~$e^{-}$ per $(1\times1)$ unit cell on average. 

The effect of the strong electron doping induced by adsorption of $0.11$~ML of K at the surface is also visible on the calculated electronic band structure in Fig.~\ref{Figure WF}~(c) along the A-L direction of the three-dimensional (3D) Brillouin zone (BZ) [see Fig.~\ref{Figure K-dep}~(a)]. The red and black band structures correspond to the pristine and K-doped IrTe$_2$ structures, respectively. The main changes occur within the first eV below the Fermi level $E_{\text{F}}$ with a shift of the surface-related electronic states towards higher BE. In particular, the band located at the A point  around $0.8$~eV below $E_{\text{F}}$ is shifted by almost $100$~meV down to high BE and is flattened. This band is known to be a surface state~\cite{RumoPRB,KingNatMat} (SS) and is consequently strongly affected by the surface doping, as discussed below.

\subsection{Electronic structure at room temperature}

\begin{figure*}[t]
\includegraphics[width=1.7\columnwidth]{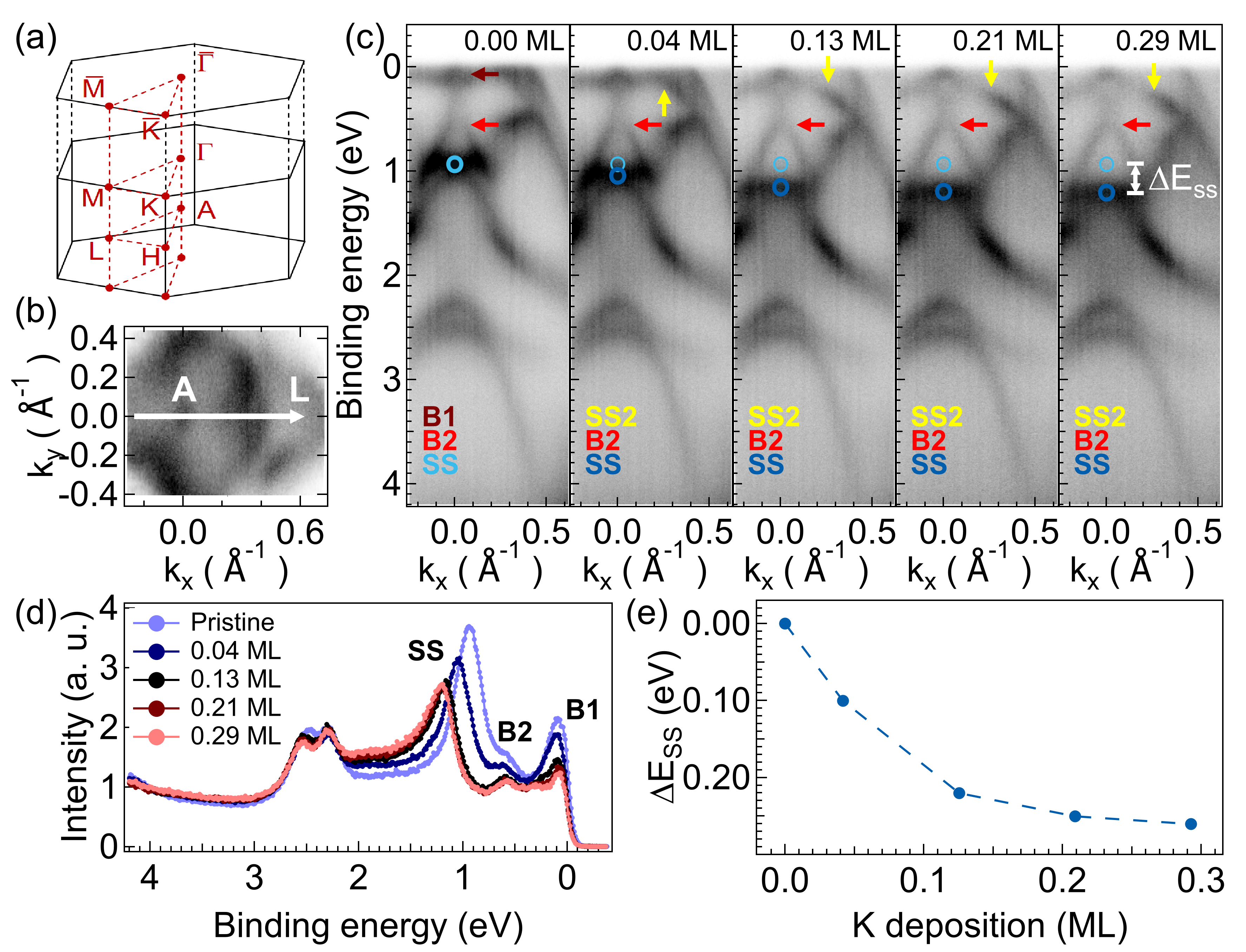}
\caption{\label{Figure K-dep} 
(Color online) (a)~The Brillouin zone of IrTe$_2$. (b)~Fermi surface of IrTe$_2$ for $h\nu = 21.22$~eV taken at $295$~K. (c)~ARPES spectra measured along AL direction for $h\nu = 21.22$~eV for a pristine, $0.04$, $0.13$, $0.21$ and $0.29$~ML K-doped crystals. (d)~EDCs for the pristine and K-doped crystals (integrated $\pm0.09$~\AA$^{-1}$ around A along AL direction). (e)~Binding energy shift of the surface state SS measured in ARPES as a function of K doping.} 
\end{figure*}

We have performed ARPES measurements as a function of K-deposition at RT to further discriminate the changes in the electronic structure of IrTe$_2$. Figure \ref{Figure K-dep}~(a) presents the 3D BZ and its surface projection. A RT Fermi surface (integrated over $0.05$~eV around $E_{\text{F}}$) is shown in Figure~\ref{Figure K-dep}~(b), obtained with a photon energy $h\nu = 21.22$~eV. At this photon energy, states close to the ALH plane are probed~\cite{OotsukiJapanLettersES,OotsukiJapanLettersPt,OotsukiJournPhys}. In Fig.~\ref{Figure K-dep}~(c), ARPES spectra taken at RT along the AL direction for a pristine crystal, as well as for $0.04$, $0.13$, $0.21$ and $0.29$~ML K-doped crystals are displayed. Corresponding energy distribution curves (EDC) integrated on a small momentum range ($\pm0.09$~\AA$^{-1}$) around A are shown in Figure~\ref{Figure K-dep}~(d). On the pristine crystal [Fig.~\ref{Figure K-dep}~(c), left panels], the electronic bands are sharp and, by comparison with the literature~\cite{LeeIOP,KingNatMat,RumoPRB}, we can identify the presence of bulk state B1 just below $E_{\text{F}}$, a second bulk band B2, and an intense surface state SS at about $1$~eV BE~\cite{RumoPRB}. With the increase of K doping, the surface state SS flattens and shifts towards higher BE from $1$ (pristine crystal) to $1.2$~eV ($0.29$~ML K-doped crystal) (also see the corresponding EDCs integrated around A in Fig.~\ref{Figure K-dep}~(d)). Note that a second surface state SS2 near $E_{\text{F}}$ exhibits a similar shift to higher BE [see Fig.~\ref{Figure K-dep}~(c)]. These observations are in line with the prediction obtained from our DFT calculations of Fig.~\ref{Figure WF}~(c), especially with respect to the surface states SS and SS2. Figure~\ref{Figure K-dep}~(e) reports the relative shift of the surface state SS ($\Delta E_{\text{SS}}$) at A as a function of K doping [see also the blue markers in Figure~\ref{Figure K-dep}~(c)], which follows the same trend as the work function [see Fig.~\ref{Figure WF}~(a)].

Overall, the K adsorption at the surface of metallic IrTe$_2$ produces a shift of the surface states, as already observed in numerous materials~\cite{HossainNatPhys,BoylePRB,KermerPRR}. However, IrTe$_2$ is a singular material since it undergoes a series of charge-ordered phase transitions upon cooling~\cite{RumoPRB,MauererPRB,KoNatCom,LiNatCom,LeeIOP}. Consequently, a crucial point to address is the impact of K surface doping on such structural instabilities. 

\subsection{Temperature dependence in ARPES}

\begin{figure*}
\includegraphics[width=1.8\columnwidth]{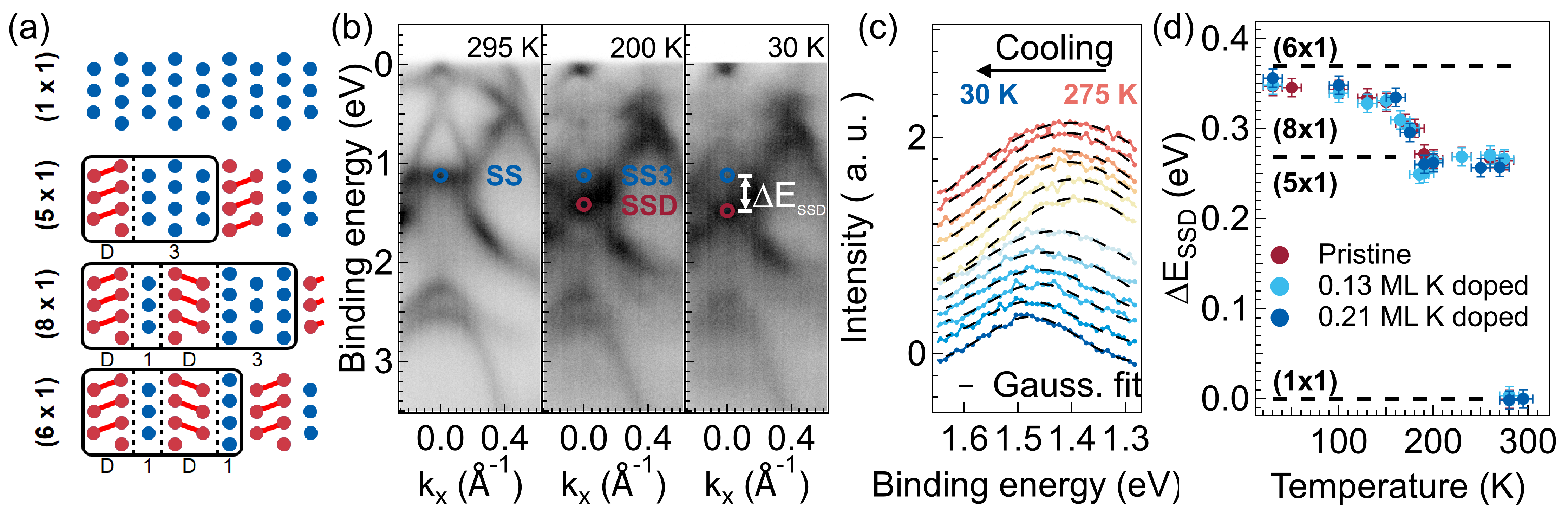}
\caption{\label{Figure Temp} 
(Color online)~(a)~Structural models of the Ir atomic planes for the different charge-ordered phases. (b)~ARPES spectra measured along AL direction for $h\nu = 21.22$~eV at three different temperatures for $0.13$~ML K~doped crystal of IrTe$_2$. (c)~Temperature-dependent EDCs upon cooling (integrated $\pm0.09$~\AA$^{-1}$ around A along AL direction). (d)~RT relative binding energy shift of the surface state SSD measured in ARPES as function of temperature for a pristine crystal, $0.13$ and $0.21$~ML K-doped crystals.} 
\end{figure*}

We have thus performed temperature-dependent ARPES measurements with a particular focus on the energy position of the surface state at around 1 eV BE that can be used as a marker of the phase transitions~\cite{RumoPRB}, since its BE depends strongly on the Ir-Ir dimer length in the stripe phases. In Fig.~\ref{Figure Temp}~(a), we recall schematically the structure of the basic building blocks for the Ir atoms in the different phases in IrTe$_2$. The ($1\times1$) phase is composed only of equivalent atoms, leading to the surface state SS at about $1$~eV BE. In the ($5\times1$) phase, five atoms split into two dimerized Ir atoms (one dimer) and three undimerized atoms [Fig.~\ref{Figure Temp}~(a)]. This leads to a splitting of the surface state SS into a contribution due to the dimerized Ir atoms, named SSD, at higher binding energy, and a contribution due to the three undimerized Ir atoms, named SS3. In the ($8\times1$) and in the $(6\times1)$ phases [Fig.~\ref{Figure Temp}~(a)], the sequence of Ir atoms changes further, implying a different combination of dimerized and undimerized atoms. At the same time, the Ir-Ir dimer length has been shown to reduce further across these phase transitions~\cite{PascutPRB,HsuPRL}.

Figure~\ref{Figure Temp}~(b) displays ARPES spectra taken at $295$~K~$>$~$T_{c_1}$, $T_{c_1}$~$>$~$200$~K~$>$~$T_{c_2}$ and $30$~K~$<T_{c_2}$, along AL direction for a K coverage of $0.13$~ML (see all ARPES data in the Appendix~\ref{Extend Temperature dependent data}). At $295$ K, sharp electronic states are clearly observed on the left panel. The surface state SS is easily distinguishable at a BE of $1.14$~eV. For temperatures below $T_{c1}$, the electronic states become more intricate due to the appearance of new translational symmetry of the charge-ordered phases of mixed orientations~\cite{HsuPRL}. A multitude of folded bands can be identified, especially in the BE range between $E_{\text{F}}$ and $2.0$~eV, see Fig.~\ref{Figure Temp}~(b). At $200$~K, in the ($5\times1$) phase, the surface state SS is split into two states, namely SS3 and SSD, the BE of the latter being now 1.4 eV. At $30$~K, in the ($6\times1$)-dominated phase [see right panels in Fig.~\ref{Figure Temp}~(b)], SSD shifts further to higher BE reaching $1.49$~eV. Figure~\ref{Figure Temp}~(c) shows EDCs at the energy of the surface state integrated on a small momentum range ($\pm0.09$~\AA$^{-1}$) around A together with the corresponding fits, consisting of a single Gauss function, for temperatures down to $30$~K. The resulting shift of the surface state SSD in BE, $\Delta E_{\text{SSD}}$, is defined as the difference between the surface state SS RT value for a given doping and the surface state SSD value at a temperature $T$ for the same doping [see also the colored markers in Fig.~\ref{Figure Temp}~(b)]. It is displayed for the pristine crystal (same data as in Ref.~\cite{RumoPRB}) as well as for the present $0.13$~ML K-doped crystal and for an additional doping of $0.21$~ML as a function of temperature in Fig.~\ref{Figure Temp}~(d). It allows us to compare directly the evolution of SSD with and without potassium.

Interestingly, the magnitudes of $\Delta E_{\text{SSD}}$ upon cooling through all three first-order phase transitions ($278$, $180$ and $165$~K) for the $0.13$~ML K-doped crystal are not only nearly indistinguishable from pristine IrTe$_2$, but also from a more K-doped sample ($0.21$~ML). This indicates that the phase transitions are remarkably insensitive to K deposition at these values of K doping.
\\

\subsection{Temperature dependence in LEED}

\begin{figure}[b]
\includegraphics[width=1\columnwidth]{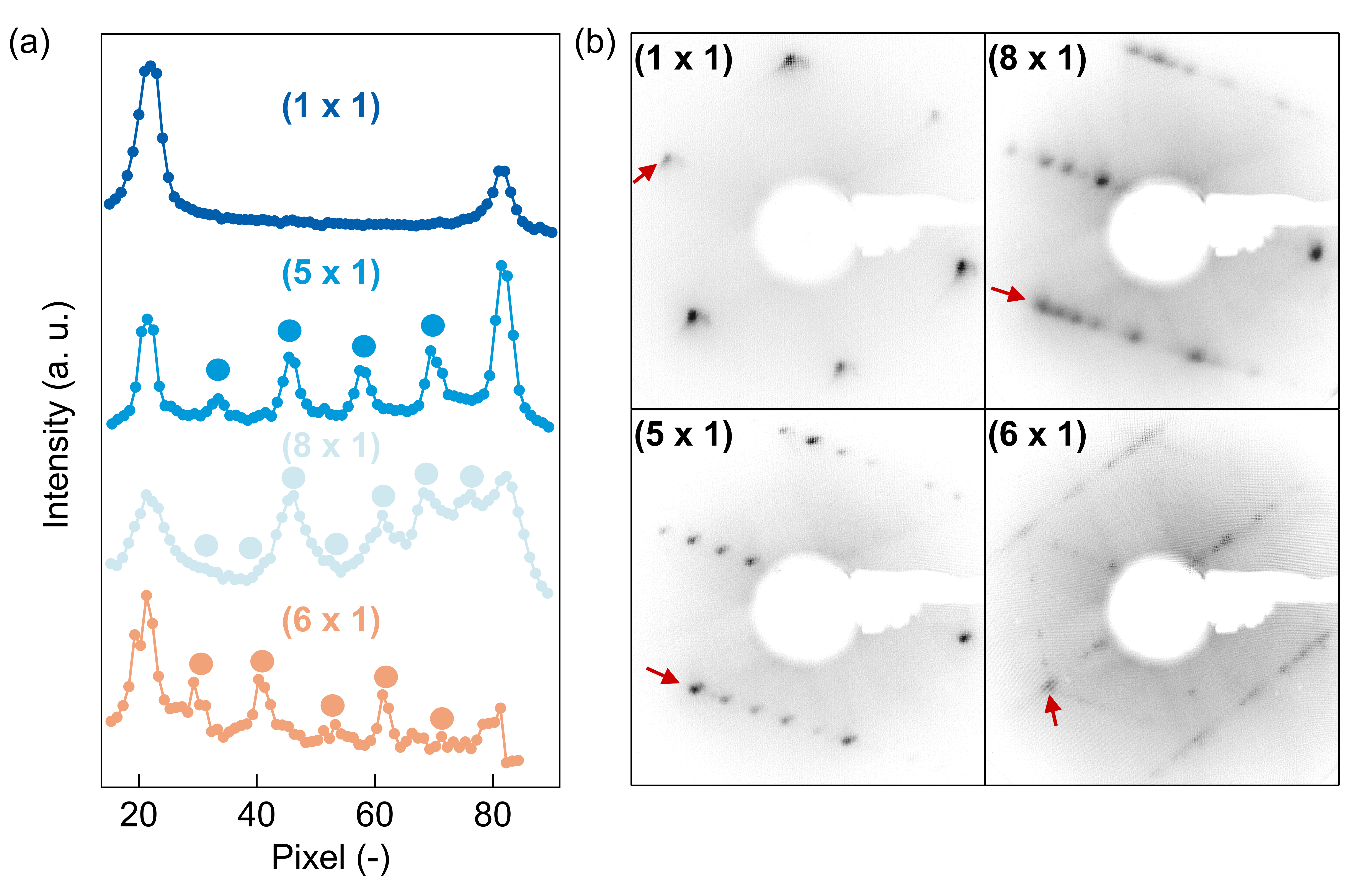}
\caption{\label{Figure K-IrTe2 LEED} 
(Color online) (a)~Line profiles of LEED images shown in (b). (b)~Raw LEED images of a $0.13$~ML~K doped IrTe$_2$ crystal in the ($1\times1$) phase at $295$~K, ($5\times1$) phase at $230$~K, ($8\times1$) phase at $170$~K and ($6\times1$) phase at $30$~K. All images were obtained using $64$~eV electron energy.} 
\end{figure}

To support the conclusions drawn from ARPES, we have also checked the periodicities of the stripe phases of the $0.13$~ML K-doped crystal of IrTe$_2$ with LEED. Figure~\ref{Figure K-IrTe2 LEED}~(a) displays the line profiles of LEED images taken at different temperatures (upon cooling) specific to the different phases of IrTe$_2$, see Fig.~\ref{Figure K-IrTe2 LEED}~(b). They confirm the presence of the ($1\times1$) phase at $295$~K, the ($5\times1$) phase at $230$~K, the ($8\times1$) phase at $170$~K and the ($6\times1$) phase at $30$~K. Note that due to the unavoidable mixture of domains of different orientations~\cite{ChenPhysRev,MachidaPRB}, lines of superstructure spots are observed in the three-fold symmetry equivalent directions (see in particular the LEED image at 30~K for the ($6\times1$) phase). Furthermore, below $T_{c_3}$, we observe coexistence of domains of the ($8\times1$) phase and of the ($6\times1$) phase (see Appendix~\ref{Extend Temperature dependent data}), a pattern already observed for pristine crystals~\cite{ChenPhysRev,RumoPRB}. Our complementary LEED study therefore confirms that the deposition of K atoms does not change the periodicities of the stripe phases observed across the phase transitions at the surface of IrTe$_2$.

\section{Discussion}

In this work, we report that K doping shifts the surface states SS at RT to higher BE, without inducing surface reconstruction (see LEED image at RT Fig.~\ref{Figure K-IrTe2 LEED}). This energy shift is due to a decrease of the surface potential, as observed for instance on the Au(111) surface~\cite{Forster_SurfSci2006}, and is therefore not related to any dimerization. This is further confirmed by the absence of the split surface state SSD at RT (see Fig.~\ref{Figure K-dep}). However, the energy shift of the surface state of the dimerized atoms, SSD, at low temperature across the phase transitions (see Fig.~\ref{Figure Temp}) does not vary with the K coverage used in this work. As motivated in our previous work~\cite{RumoPRB}, this means that the dimer length does not change and that the charge-ordered phase transitions are not modified by the surface K doping.

In contrast, many layered materials turn out to be very sensitive to alkali doping. For instance, the CDW phase in the TMDC TiSe$_2$ is already completely suppressed at about 0.1 ML of Rb coverage~\cite{RossnagelIOPTiSe}. The band gap of the semiconducting layered-material black phosphorous is very sensitive to alkali doping, already decreasing above $0.1$~ML and this material has been shown to undergo a semiconductor-semimetal transition at $0.35$~ML~\cite{KimScience}. The same effect has been discovered also in Ta$_2$NiSe$_5$~\cite{FukutaniPRL}, for which a semiconductor-semimetal transition occurs at about $0.15$~ML of K doping~\cite{ChenPRB}. These recent results on low-dimensional layered materials emphasize the surprising insensitivity of IrTe$_2$ to K doping at low coverages.

What is then the reason for this robustness of the surface transitions in IrTe$_2$? IrTe$_2$ has a metallic behavior above and below the phase transitions, with a significant density of states maintained at the Fermi level~\cite{FangScienRep,PascutPRL}, in contrast with the materials mentioned above. This is due to the fact that a dramatic reduction of density of states is observed only for the bonding and antibonding dimer states that are not lying close to $E_F$~\cite{SalehEntropy,PascutPRL,IdetaSciAdv,KimPRL}. Away from these dimer states the density of states is insignificantly affected by the phase transitions~\cite{PascutPRL}. This naturally makes the charge-ordered phase transitions in IrTe$_2$ insensitive to carrier doping, for three reasons. (1) Screening of the electric field of the ionized doping atoms is more efficient for a metallic surface. This excludes the possibility of a large unscreened electric field near the surface, which is central to the phase transition in black phosphorus for instance~\cite{KimScience}. (2) When electron doping changes the low-energy band structure, and thus the corresponding Fermi surface shape, it can significantly alter the electronic susceptibility of materials, reducing their tendency to instabilities, as for instance in TiSe$_2$~\cite{jaouenarxiv} or Ta$_2$NiSe$_5$~\cite{FukutaniPRL}. This mechanism is often the cause of the suppression of a collective instability, like a charge density wave, by alkali doping. However, the low-energy electronic structure of IrTe$_2$ is little modified by a given electron doping density in a rigid band model picture, since the high density of states at $E_F$ can easily accommodate the doping electrons. (3) Finally, the doping charge does not populate the antibonding states that are further away from $E_F$, since the nonbonding states near $E_F$ likely capture the K doping electron. This prevents the destabilization of Ir dimers. Furthermore, we recall here that our DFT calculations indicate that the charge transfer from the K adatom affects only the top surface layer and mainly dopes the surface states, a fact confirmed by our ARPES data, so that the bulk electronic structure stays unchanged. We further emphasize that, despite this surface doping, it is surprising that the $(6\times1)$ phase transition, which occurs only at the surface, is not affected. It demonstrates that purely electronic doping is not the most efficient way to destabilize the charge-ordered phases of IrTe$_2$. Therefore, our results support the idea that the instability is driven by a local mechanism mainly involving the atoms forming the dimers, stabilized by a large electronic energy gain due to the creation of the bonding-antibonding states of these dimers~\cite{PascutPRL,PascutPRB,SalehEntropy}. 

In contrast, recent literature reveals that structural perturbation, as induced by chemical substitution~\cite{PyonJPSJ,OhPRL,YangNatPhys} or applied strain~\cite{NicholsonComMater2021}, is much more efficient at altering the phase diagram. More specifically for $4\%$ of Pt or Pd substitution, which kills the charge order in favor of superconductivity, a gain of $0.04$~$e^{-}$ per unit cell throughout the crystal is achieved, while in our case $0.07$~$e^{-}$ per unit cell localized at the surface is observed from our calculations. However, the chemical substitution causes a local strain in the lattice. For example, Pd has an atomic size difference of $1\%$ with Te, which creates distortions in the lattice, which itself contributes to the phase transitions. This is much greater than the $0.1\%$ mechanical strain which is observed to stabilise the $(6\times1)$ phase. The phase transitions are strongly disturbed in these particular cases while pure electronic doping is unable to induce such an effect. This is consistent with the results of a recent time-resolved ARPES study~\cite{MonneyPRB} that concluded that infrared photoexcitation cannot efficiently trigger the phase transition and that only a partial photoinduced phase transition occurs, driven by the transient heating of the lattice.

\section{Conclusion}

In this work, we have studied the influence of K-doping on IrTe$_2$ with ARPES and LEED, supported by DFT calculations. We have performed a detailed and systematic analysis as a function of K deposition at RT and we have probed the occurrence of charge-ordered phases as a function of temperature. We have shown that the perturbation by electronic doping from the alkali atoms has no effect on the charge-ordered phase transitions at the surface of IrTe$_2$. This emphasizes the important role of the structural component in the stabilization of these phases. In this framework, it will be particularly interesting to test the stability of the charge-ordered phases of IrTe$_2$ in the limit of a single monolayer or within of van der Waals heterostructures.

\section{Acknowledgments} 

This project was supported from the Swiss National Science Foundation (SNSF) Grant No. P00P2\_170597. A.P. acknowledges the Osk. Huttunen Foundation for financial support, and CSC – IT Center for Science, Finland, for computational resources. We are very grateful to P.~Aebi for fruitful discussions and for sharing with us his photoemission setup. Skillful technical assistance was provided by F.~Bourqui, B.~Hediger, J.L.~Andrey, O.~Raetzo and M.~Andrey. 

\newpage
\section{Appendices}

\subsection{K deposition calibration}
\label{Potassium deposition calibration}

\begin{figure}[h!]
\includegraphics[width=0.7\columnwidth]{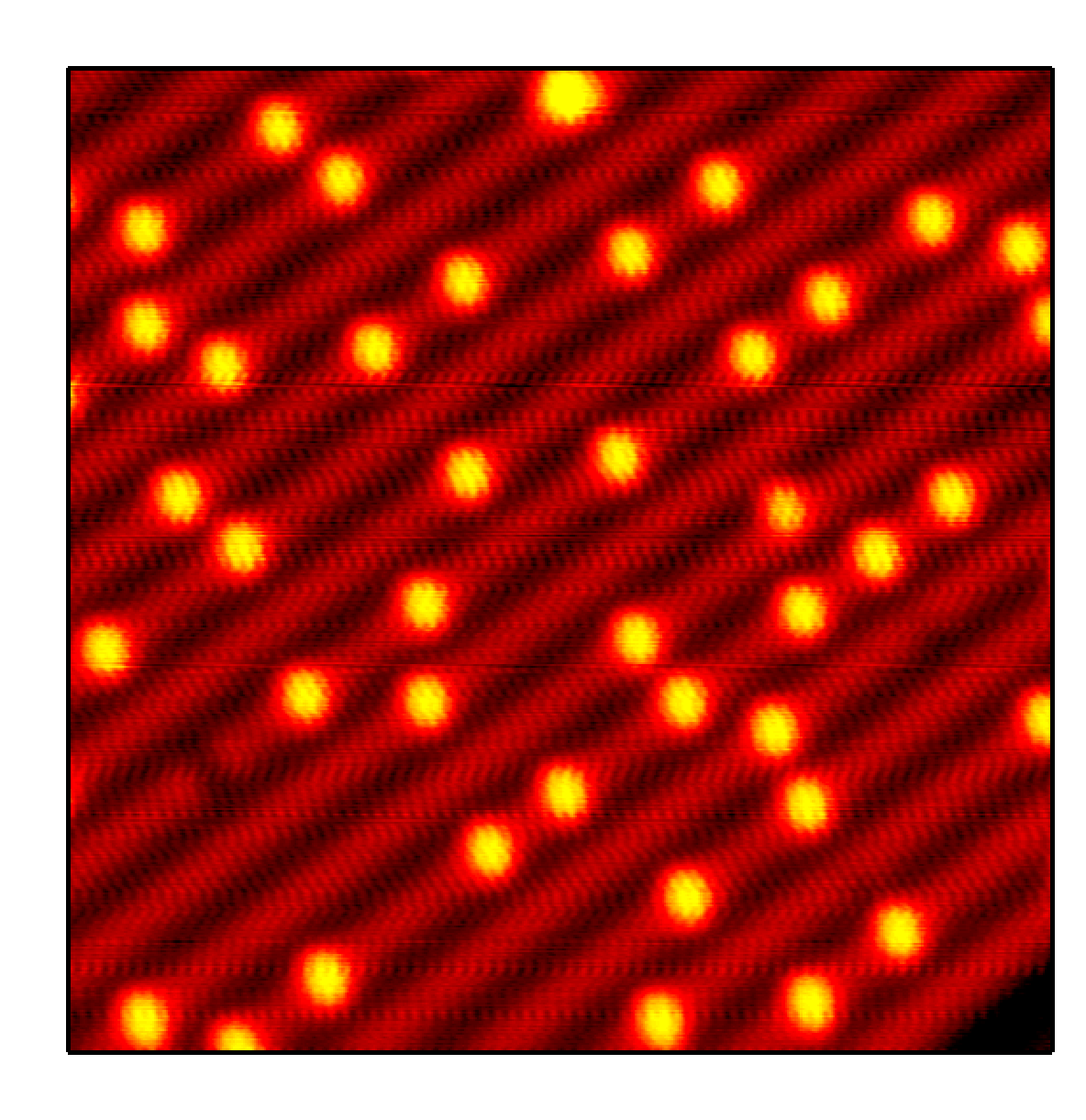}
\caption{\label{Figure STM} 
(Color online)~(a) $19.7\times19.7$~nm$^2$ constant current mode STM image of a $0.01$~ML K-doped IrTe$_2$ sample, $V_{\text{bias}} = -200$~mV, $I = 0.2$~nA.} 
\end{figure}

Figure \ref{Figure STM} shows a high-resolution STM image taken on a K-doped IrTe$_2$ crystal of the occupied states at a bias voltage of $V_{\text{bias}}=-200$ mV and a constant current mode of $I = 0.2$~nA with atomic resolution. The K deposition was performed at RT. The STM image [Fig.~\ref{Figure STM}] reveals the presence of adsorbed K atoms at the surface as bright protrusions on top of the low temperature reconstruction of IrTe$_2$. From the statistics made on a few images, a K coverage of $0.01$~ML is estimated for an exposition time to the SAES getter source of $30$~s (one K atom per IrTe$_2$ surface unit cell at RT is defined as $1$~ML of K atom). 

We caution that this STM measurement reveals only adsorbed K atoms at the surface of IrTe$_2$. However, intercalation of K atoms in the van der Waals gap could also occur in parallel to adsorption, as reported in the literature for TaS$_2$ or graphite~\cite{RossnagelIOPTiSe,CaragiuJPhys,RamirezPRB}. We stress that the distance between two sandwiches of IrTe$_2$ (interlayer spacing) is smaller (about $2.7$ \AA) than for most TMDCs and for graphite ($3.3$~\AA). Therefore we expect alkali intercalation to be  unfavorable for the present case.

\subsection{Supplementary data for the temperature dependent study}
\label{Extend Temperature dependent data}

\begin{figure}[h!]
\centering
\includegraphics[width=1\columnwidth]{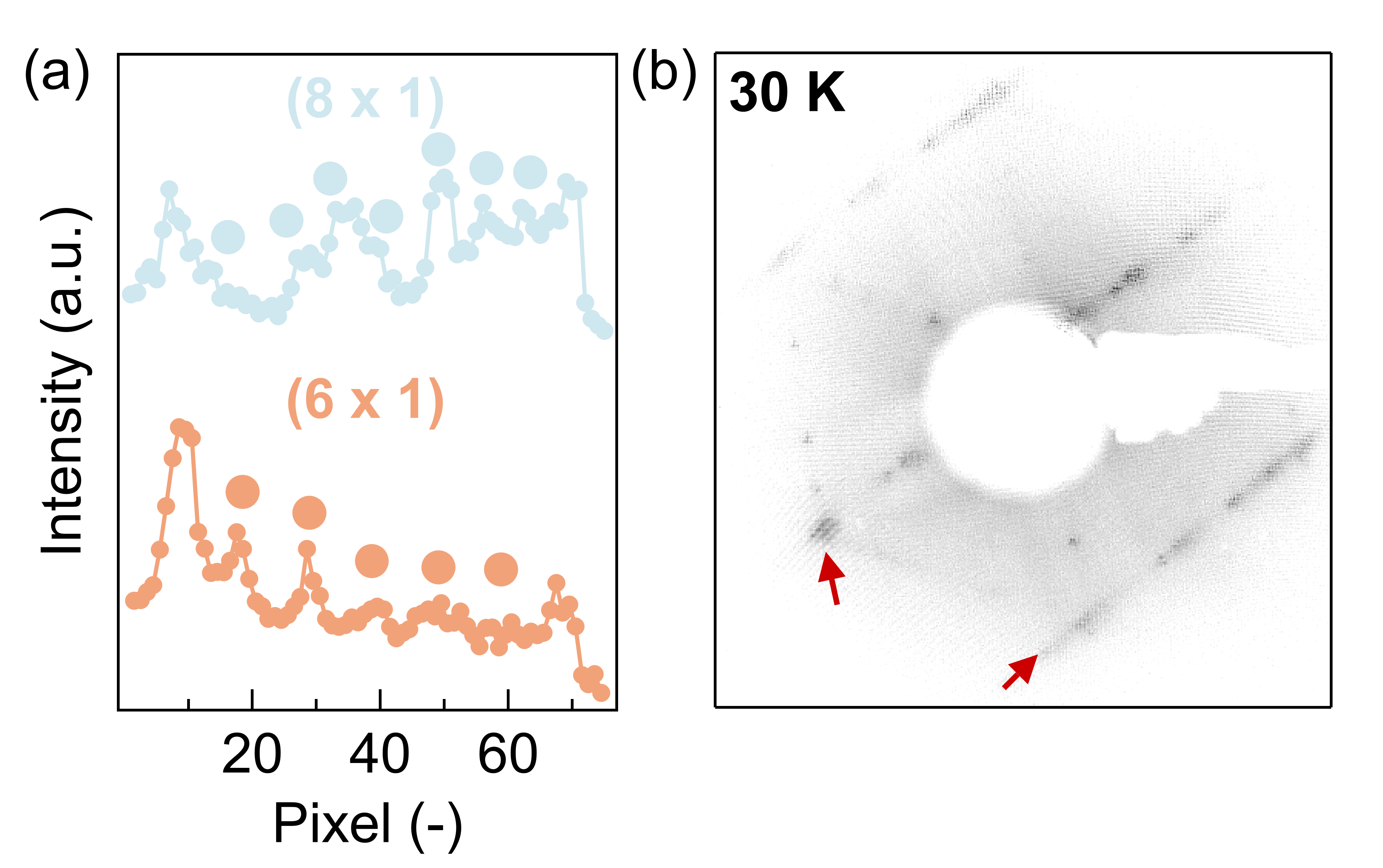}
\caption{(a)~Lines profiles of LEED images shown in (b). (b)~LEED image from a $0.13$~ML~K doped crystal of IrTe$_2$ in the ($8\times1$) and ($6\times1$) phases at $30$~K. All images were obtained using $64$~eV electron energy.\label{Figure LEED MixPhase}} 
\end{figure}

Figure~\ref{Figure LEED MixPhase} displays a LEED image of the $0.13$~ML K doped IrTe$_2$ showing both the ($8\times1$) and the ($6\times1$) phases at $30$~K, as observed in the pristine IrTe$_2$ case for the same temperature.
\\

\begin{figure}[t]
\centering
\includegraphics[width=0.9\columnwidth]{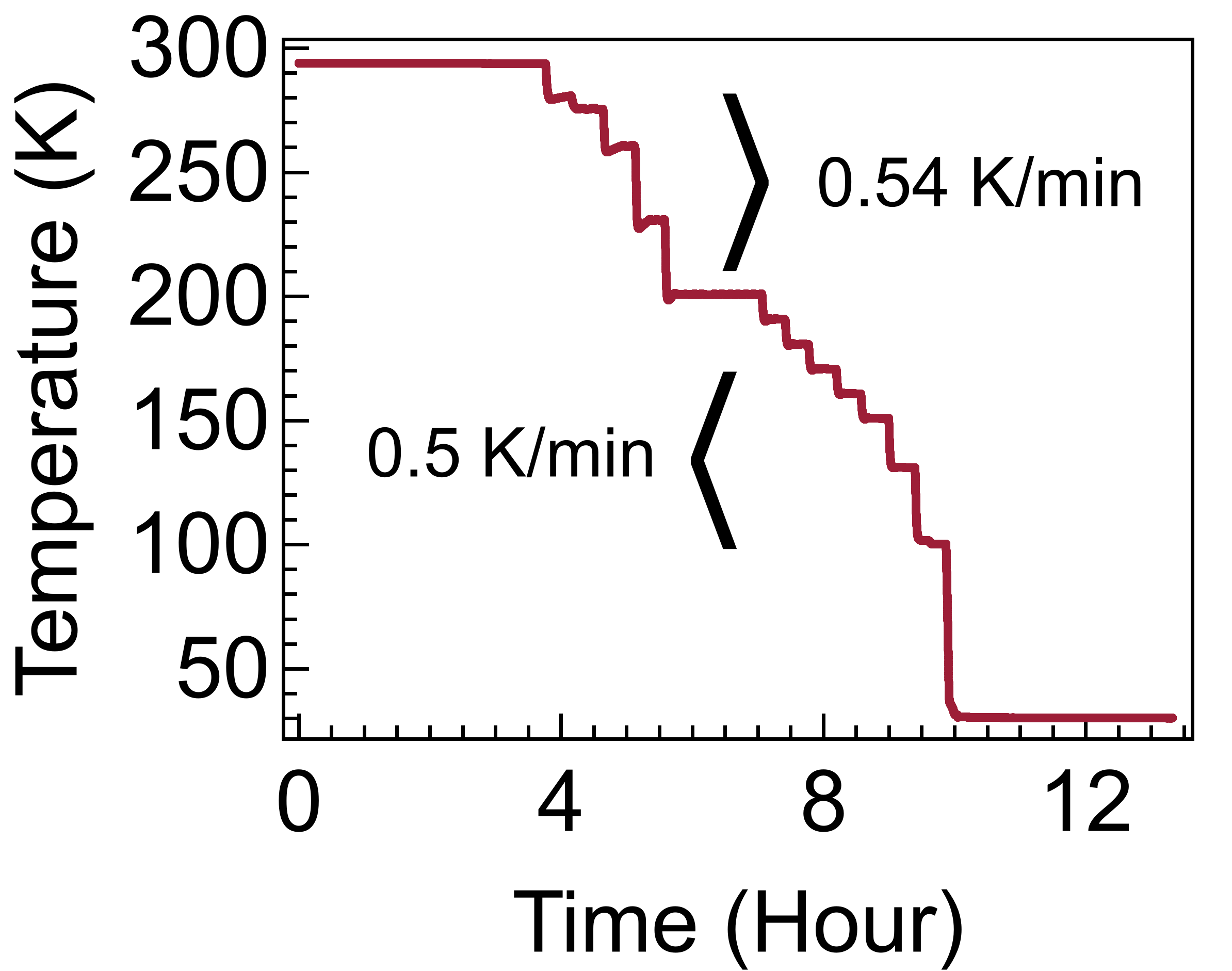}
\caption{Temperature of the sample as a function of time during the temperature dependence study.\label{Figure Temperature ARPES}} 
\end{figure}

Figure~\ref{Figure Temperature ARPES} displays the evolution of the temperature of the sample during the temperature dependent ARPES measurements. The cooling rate is less than 3 K/min and on average less than $0.55$~K/min.
\\

\begin{figure*}
\centering
\includegraphics[width=1.9\columnwidth]{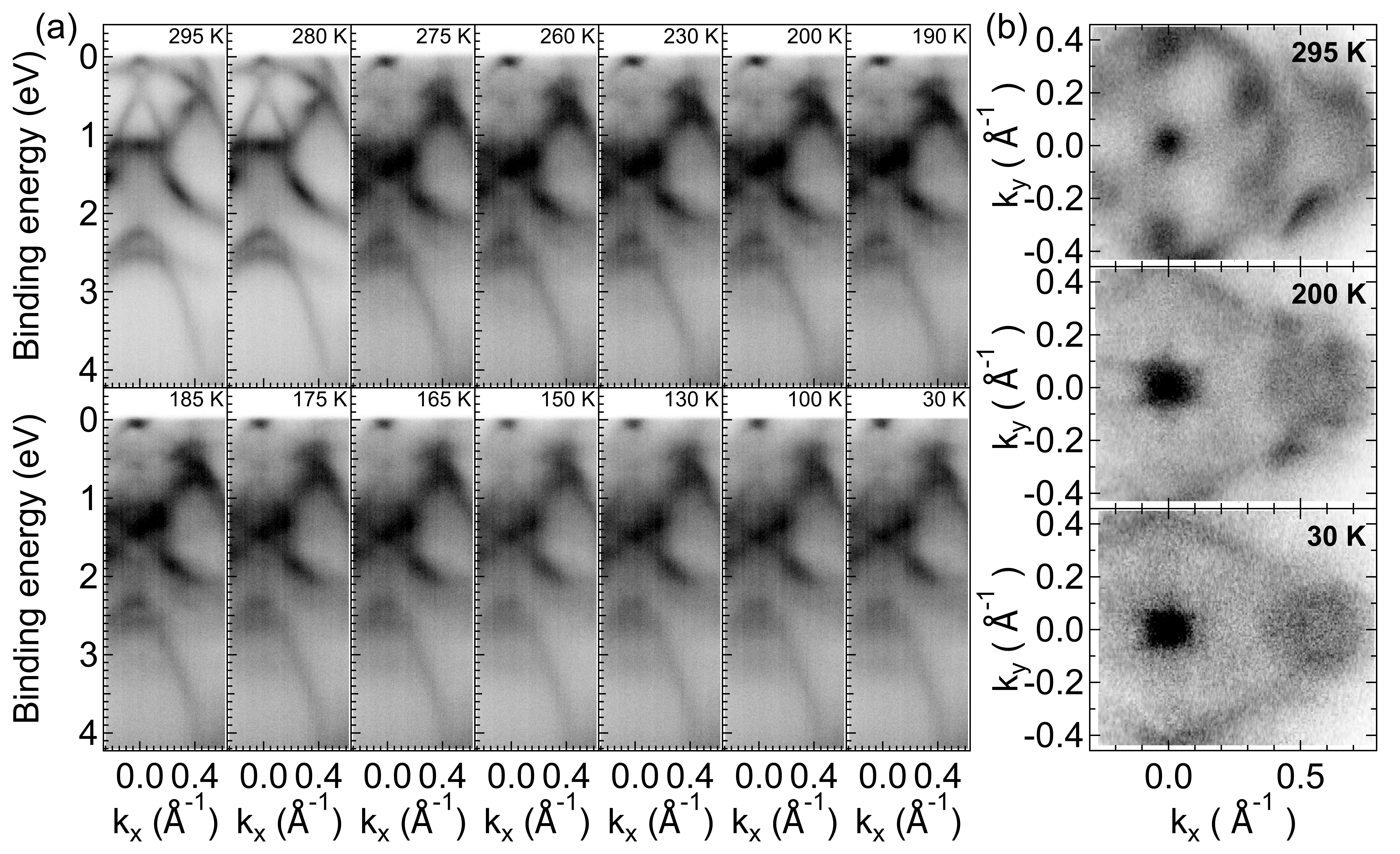}
\caption{(Color online)~(a) ARPES spectra of a $0.13$~ML K-doped IrTe$_2$ crystal measured along AL direction with a photon energy of $h\nu=21.22$~eV upon cooling.(b) Fermi surfaces of a $0.13$~ML K-doped IrTe$_2$ crystal integrated over $0.05$~eV around $E_{F}$ at different temperatures upon cooling.\label{Figure ARPES supp}} 
\end{figure*}

Figure~\ref{Figure ARPES supp}~(a) shows ARPES spectra taken at different temperatures during the cooling process, $295$, $280$, $275$, $260$, $230$, $200$, $190$, $185$, $175$, $165$, $150$, $130$, $100$ and $30$~K on a $0.13$~ML K-doped crystal of IrTe$_2$. The evolution of the electronic structure of a $0.13$~ML K-doped crystal of IrTe$_2$ can be observed and in particular the shift of the surface state as a function of temperature. Figure~\ref{Figure ARPES supp}~(b) displays Fermi surfaces of a $0.13$~ML K-doped IrTe$_2$ crystal at different temperatures.

%%%\bibliography{library1}

\end{document}